\documentstyle[prl,aps,multicol,epsfig]{revtex} 
\draft
\title{Critical Casimir effect and wetting by helium mixtures} 
\author{T. Ueno\footnote{Present address: Research Laboratory
of Electronics and MIT-Harvard Center for Ultracold Atoms,
Massachusetts Institute of Technology, Cambridge, Massachusetts 02139, USA}, 
S. Balibar, T. Mizusaki\footnote{Permanent address: Department
of Physics, Graduate School of Science, Kyoto University,
Kitashirakawa-Oiwake-cho, Sakyo-ku, Kyoto 606-8502, Japan}, F. Caupin
 and E. Rolley}
\address{Laboratoire de Physique Statistique de l'Ecole Normale Sup\'erieure \\
 associ\'e aux Universit\'es Paris 6 et Paris 7 et au CNRS \\
 24 rue Lhomond 75231 Paris Cedex 05, France\\}
\begin{document}
\maketitle
\begin{abstract}
We have measured the contact angle of the interface of phase-separated
$^{3}$He-$^{4}$He mixtures against a sapphire window.  We have found
that this angle is finite and does not tend to zero when the
temperature approaches $T_t$, the temperature of the tri-critical
point.  On the contrary, it increases with temperature.  This behavior
is a remarkable exception to what is generally observed near critical
points, i.e. ``critical point wetting''.  We propose that it is a
consequence of the ``critical Casimir effect'' which leads to an
effective attraction of the $^{3}$He-$^{4}$He interface by the
sapphire near $T_{t}$.
\end{abstract}
\begin{multicols}{2}

In 1977, J. W. Cahn predicted that ``in any two-phase mixture of
fluids near their critical point, contact angles against any third
phase become zero in that one of the critical phases completely wets
the third phase and excludes contact with the other critical
phase''~\cite{cahn}.  This ``critical point wetting'' is a very
general phenomenon~\cite{cahn,heady,indekeu,bonn}. We 
found an exception to it by studying helium mixtures in
contact with a sapphire window~\cite{exceptions}.

\begin{figure}
\centerline{\includegraphics[width=1\linewidth]{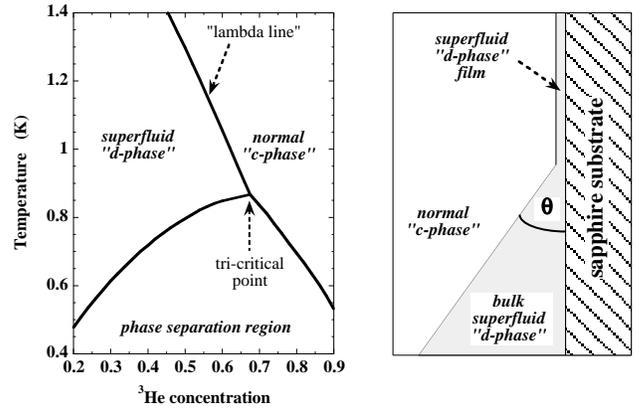}}
\caption{The phase diagram of $^{3}$He-$^{4}$He mixtures (left graph). 
On the right, a schematic view of the contact angle $\theta$. There is a 
superfluid film of $^{4}$He rich ``d-phase''  between the substrate and the c-phase. Its  
thickness being finite, $\theta$ is non-zero.}
\label{fig:contactangle}
\end{figure}

In fact, de Gennes~\cite{pgg} had noticed that long range forces may prevent
complete wetting.  Nightingale and Indekeu~\cite{night} further
explained that if a long range attraction is exerted by the third
phase on the interface between the two critical phases, partial
wetting may be observed up to the critical point.  We propose that, in
$^{3}$He-$^{4}$He mixtures near their tri-critical point, this
attraction is provided by the confinement of the fluctuations of
superfluidity, i.e. a critical Casimir
effect~\cite{pgg2,night,krech,garcia,garcia2} in the $^{4}$He-rich
film between the sapphire and the $^{3}$He-rich bulk
phase~(Fig.~\ref{fig:contactangle}).
 
For a solid substrate in contact with a phase-separated 
$^{3}$He-$^{4}$He mixture, complete wetting by the $^{4}$He-rich
``d-phase'' was generally expected, due to the van der Waals
attraction by the substrate~\cite{romagnan,sornette}.  However, we
measured the contact angle $\theta$ of the $^{3}$He-$^{4}$He interface on
sapphire, and we found that it is finite. Furthermore, it increases 
between 0.81 and 0.86~K, close to the tri-critical
point at $T_{t}$~=~0.87~K~\cite{jltp}.  This behavior is opposite to
the usual ``critical point wetting'' where $\theta$ decreases
to zero at a wetting temperature $T_{w}$ below the critical point.  In
this letter, we briefly recall our experimental results before
explaining why the ``critical Casimir effect'' provides a reasonable
interpretation of our observations.
We use a dilution refrigerator with optical access~\cite{jltp}.
Our liquid sample is at saturated vapor pressure, 
and confined between two sapphire windows which form an interferometric cavity. 
The inside of the windows is treated to have a 15$\%$
reflectivity. The cell is made of pure copper and 
neither the windows nor the helium absorb any light, so that a very 
good thermal homogeneity is achieved. From fringe patterns, we analyze the
profile of the c-d interface near its contact line with one of the
windows~\cite{jltp,etienne}.  A fit with a solution of Laplace's
equation gives the interfacial tension $\sigma_{i}$ and the contact
angle $\theta$.  As $T$ approaches $T_{t}$, the
capillary length vanishes so that the region to be analyzed becomes
very small.  However, our typical resolution is 5~$\rm \mu m$,
significantly smaller than the capillary length (from 84~$\rm \mu m$
at 0.81~K to 33~$\rm \mu m$ at 0.86~K).  Here, we present results only
in this temperature range because, closer to the tri-critical point,
we would need a better resolution and, below 0.80~K, refraction effects distort the
fringe patterns \cite{jltp}.

\begin{figure}
\centerline{\includegraphics[width=1\linewidth]{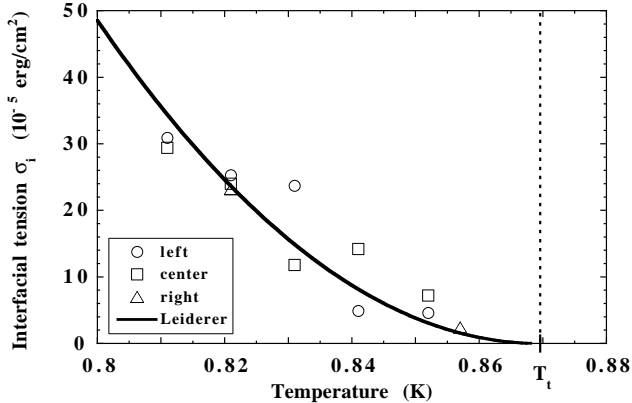}}
\caption{Our measurements of the interfacial tension 
agree with Leiderer's results (solid line). Different 
symbols correspond to three different positions along the contact line.}
\label{fig:tension}
\end{figure}
\begin{figure}
\centerline{\includegraphics[width=1\linewidth]{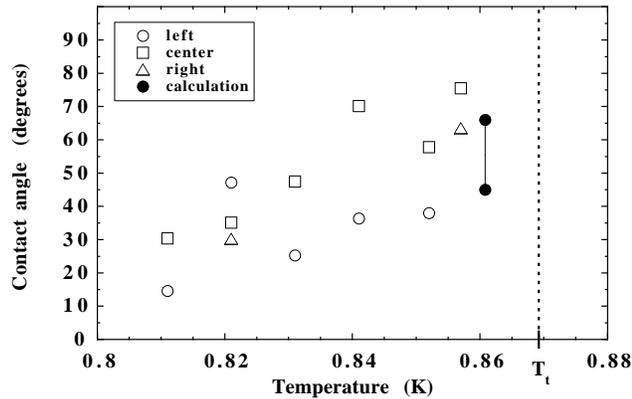}}
\caption{Temperature dependence of the contact angle $\theta$. Black 
symbols correspond to the present calculation.}
\label{fig:angle}
\end{figure}
For each temperature, we analyzed three pictures at different
positions along the contact line.  As shown in Fig.~\ref{fig:tension},
our measurements of $\sigma_{i}$ agree well with Leiderer's
result $\sigma_{i}$ = 0.076 t$^{2}$ erg/cm$^{2}$
($t=(1-T/T_{t})$ is the reduced temperature)~\cite{leiderer}.  As for the contact
angle $\theta$ (Fig.~\ref{fig:angle}), we found that it is non-zero
and that it increases with $T$.  On these measurements, the typical
error bar is $\pm$ 15 $^{\circ}$.  It originates in several
experimental difficulties such as the precise location of the
contact line and a slight bending of the windows which are under
stress~\cite{jltp}.

When cooling down a homogeneous mixture with concentration higher than
the tri-critical value $X_{t}$, J.P.~Romagnan et al.~\cite{romagnan}
found that a superfluid film formed between the bulk mixture and a
metallic substrate.  As they approached $T_{eq}$ where separation into
``c- '' and ``d-'' phases occurred, they observed a film thickness
diverging as $(T-T_{eq})^{-1/3}$.  This behaviour is
characteristic of the van der Waals attraction by the substrate, which
is stronger on the densest phase~\cite{sornette}.  One used to believe
that van der Waals forces were the only long range forces in this
problem, so that the film thickness should diverge to infinity, and
complete wetting by the superfluid d-phase should occur.  However,
Romagnan et al.  only measured this thickness up to about 20 atomic
layers (60 \AA). If other forces act on the film near the
tri-critical point, its thickness can saturate at a value larger than
60~\AA. R.~Garcia and M.~Chan\cite{garcia} have shown that superfluid
films of pure $^{4}$He get thinner near $T_{\lambda}$, due to the
critical Casimir effect.  Our situation is similar: our d-phase film
is just below its superfluid transition, and we have calculated that
Casimir forces limit the film thickness to a few hundred \AA~.

An increasing variation of $\theta (T)$ is also surprising.
Young's relation writes:
\begin{equation}
    \cos(\theta) = \frac{\delta \sigma}{\sigma_{i}}
    = \frac{\sigma_{sc}-\sigma_{sd}}{\sigma_{i}}
\label{eq:young}
\end{equation}
As the critical point is approached, both $\sigma_{i}$ and
$\delta \sigma$ tend to zero. It is often assumed that
$\delta \sigma$ is proportional to the 
difference in concentration between the two phases. If this was always true, 
the critical exponent of $\sigma_{i}$ would always be larger than that of 
$\delta \sigma$~\cite{indekeu}. Consequently, 
$\theta$ would always decrease to zero at a wetting temperature 
$T_{w}$ below the critical point. Our observations show that this 
reasoning does not apply to helium mixtures.

Let us now follow D.~Ross et al.~\cite{ross} to calculate $\theta$.  We first
calculate the ``disjoining pressure'' $\Pi(l)$ as a function of the
thickness $l$ of the d-phase film (Fig.~\ref{fig:contactangle}).
For this we consider three long range forces: the van der
Waals force, the Casimir force and the ``Helfrich'' force~\cite{helfrich}.
At the equilibrium film thickness $l \: = \: l_{eq}$, $\Pi(l)$
has to cross zero with a negative slope.  If $l_{eq}$ was
macroscopic, the substrate (s) to c-phase interface would be made of
an s-d interface plus a c-d interface.  Its energy per unit area 
would thus be
$\sigma_{sc} \:=\: (\sigma_{sd}\: +\: \sigma_{i})$.  If
$l_{eq}$ is small, a correction to the
above formula has to be added, which is the integral of the disjoining
pressure from infinity to $l_{eq}$.  Finally, Young's relation imply
\begin{equation}
    \cos(\theta) = \frac{\sigma_{sc}-\sigma_{sd}}{\sigma_{i}} 
    = 1 + \frac{ \int_{l_{eq}}^{\infty} \Pi (l) dl}{\sigma_{i}}
\label{eq:young2}
\end{equation}
Let us start with the van der Waal contribution $\Pi_{vdW}(l)$ to $\Pi (l)$.
 The net effective force on the
interface is the difference between the respective van der Waals 
attractions on
the d- and c-phase.  For helium on copper, Garcia
found that this attraction
 is $A_{0}/Vl^{3}$, with $A_{0}$ = 2600 K.\AA$^3$ ($V$ is the atomic 
 volume)~\cite{garcia}. 
For our window with its insulating coating, 
we expect a smaller value. Sabisky~\cite{sabisky} found $A_{0}$
= 980 K.\AA$^{3}$ for liquid $^{4}$He on CaF$_{2}$. We thus estimate
$A_{0}* \approx$ 1000 K.\AA$^{3}$ in our case. 
As for the coefficient of the differential force, it is now
\begin{equation}
A = A_{0}* \left (\frac{1}{V_{d}*}-\frac{1}{V_{c}*}\right )
\end{equation}
where $V_{c,d}$ are the respective atomic volumes in the two 
phases~\cite{kierstead}. We included 
the retarded term in the van der Waals potential~\cite{garcia} and 
finally found
\begin{equation}
    \Pi_{vdW}(l) = \frac{A}{l^{3}(1 + l/193)} = 
    \frac{14.72 t - 2.82 t^{2} 
    + 2.29 t^{3}}{l^{3}(1 + l/193)}
\label{eq:vdW}
\end{equation}
in K/\AA$^{3}$ with $l$ in \AA. 

Let us now consider the Casimir force. 
Following Garcia~\cite{garcia}, the  
confinement of superfluid fluctuations inside a film of thickness 
$l$ gives a contribution to the disjoining pressure
\begin{equation}
 \Pi_{Cas}*(l) = \frac{\vartheta (x)\:T_{t}}{l^{3}}
\label{eq:casimir}
\end{equation}
where $x = tl$ and the ``scaling function'' $\vartheta (x)$ is negative,
with a minimum of about -1.5 at $x \approx 10$. The sign of 
$\vartheta (x)$ 
depends on the symmetry of the boundary conditions on the two sides 
of the film~\cite{garcia2,krech}. In Ref.~\cite{garcia} as in our 
case, the whole film is superfluid except near both interfaces where 
the order parameter vanishes on a distance $\xi$, the correlation length. 
Consequently, the boundary conditions are symmetric for this order parameter 
and $\vartheta (x)$ is negative, meaning an attractive force. Note that 
 in Garcia's second experiment~\cite{garcia2} on mixtures, the film 
 was separated into a superfluid subfilm near the wall and a normal 
 one near the liquid-gas interface; 
 Garcia considered this as an anti-symmetric situation leading to a 
 repulsive force between the wall and the liquid-gas interface. 
 This second experiment is different from ours because  
 it measures a Casimir force on a liquid-gas surface while ours has to 
 do with the c-d interface. As a result our experiment 
 is paradoxically more similar to Garcia's first 
 experiment~\cite{garcia} with pure $^{4}$He which has also symmetric 
 boundary conditions than with Garcia's second experiment on 
 mixtures~\cite{garcia2}.
 
\begin{figure}
\centerline{\includegraphics[width=1.1\linewidth]{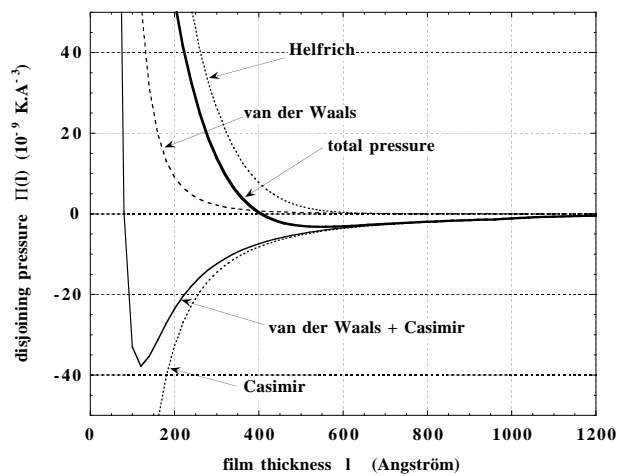}}
\caption{The three contributions to the disjoining pressure 
 for $t$ = 10$^{-2}$. The total pressure is zero at $l_{eq}$ = 400 \AA.}
\label{fig:casimir}
\end{figure}

In order to evaluate $\vartheta (x)$, and in the absence of any other 
determination, we have taken
Garcia's curve labelled ``Cap.~1'' in Ref.~\cite{garcia}. It corresponds to a 
film thickness of about 400 \AA~, as  
found below for $t$ = 10$^{-2} $.
Fig.~\ref{fig:casimir} shows that, at this temperature, the resulting Casimir contribution 
dominates the van der Waals one above about 100 \AA. This is because 
the coefficient $A$ in Eq.~4 vanishes with $t$, so that, for $t$ = 
$10^{-2}$, it is about 0.15, ten times 
less than the maximum amplitude of $\vartheta (x)$.
We still need to discuss our approximations further. 
We are dealing with a 
tri-critical point instead of the lambda transition in Garcia's 
case~\cite{garcia}. According to 
Krech and Dietrich~\cite{krech} the Casimir amplitude
is twice as large for tri-critical 
points compared to ordinary critical points. Doubling Garcia's scaling 
function enlarges $\theta$ and improves the agreement with our 
experiment (see below). Furthermore, in our system, 
concentration and superfluidity fluctuations are coupled together. 
Both should be considered in a rigorous calculation which has not yet 
been done. The boundary conditions are symmetric for superfluidity but 
they are anti-symmetric for 
concentration fluctuations since the film is richer in $^{4}$He near the 
substrate than near the c-phase. We thus believe that a rigorous 
calculation should include two contributions with opposite sign. We assume 
that the confinement of superfluidity dominates
because the Casimir amplitude is roughly proportional to the 
dimension $N$ of the order parameter~\cite{krech} ($N$ = 2 for 
superfluidity and $N$ = 1 for concentration). We hope that our 
intuition can be confirmed by further theoretical work.
The third contribution $\Pi_{H}* (l)$ originates in the limitation of
the amplitude $z$ of the c-d interface fluctuations to 
a fraction of the thickness $l$. According to Helfrich\cite{helfrich}, $<(z)^{2}> \approx 
l^{2}/6$ and, following Ross et al~\cite{ross}
\begin{equation}
    \Pi_{H} (l) = \frac{T}{2L^{2}*l}\; ,
\end{equation}
where $L$ is a long wavelength cutoff. L can be calculated from the equipartition 
theorem as
\begin{equation}
   L*=*\xi \exp{\left(\frac{2\pi\sigma_{i}* l^{2}*}{6k_{B}*T}\right )}
\end{equation}
The bulk correlation length $\xi$ is related to the surface tension $\sigma_{i}*$
 by $\xi^{2}*\approx k_{B}*T/(3\pi\sigma_{i})$, 
where the factor 3 is consistent with both Refs.~\cite{leiderer} and 
\cite{ross}. Finally 
\begin{equation}
   \Pi_{H}(l) = \frac{3\pi \sigma_{i}*}{2l} \exp{\left ( 
   \frac{-2\pi\sigma_{i}* l^{2}*}{3k_{B}*T}\right )}
\end{equation}

The disjoining pressure and the equilibrium film thickness are now
obtained by adding the three above contributions and by looking for
$l_{eq}$ such that $\Pi (l_{eq})$ = 0.  Fig.~\ref{fig:casimir} shows
the results of a calculation for $t$ = 10$^{-2}$.  If we had the van
der Waals contribution only, the disjoining pressure would be positive
everywhere and it would repell the film surface to infinity (complete
wetting).  The Casimir contribution is negative and large enough to 
induce partial wetting.  As for the Helfrich repulsion, it is very large
at small thickness but it decreases exponentially so that its effect
is to shift the equilibrium thickness by a few hundred \AA.
Fig.~\ref{fig:casimir} shows that, for $t$ = 10$^{-2}$, $l_{eq}$ = 400
\AA. This is larger than $\xi\approx $ 100 \AA,  so that the superfluidity is well 
established in the middle of the d-phase film. At this temperature
we finally calculated the contact angle with
Eq.~\ref{eq:young2}, and found $\theta$ = 45 degrees, in good
agreement with experimental results (Fig.~\ref{fig:angle}). In order
to account for tri-criticality, one could double the Casimir
amplitude; this would roughly double (1- $\cos \theta$) and change 45
into 66 degrees, in even better agreement with our data.

The most important result is that $\theta$ is finite.  Its exact
magnitude depends on the many approximations made above,
especially on the value of $\vartheta (x)$ which is only
known through Garcia's measurement in a
slightly different situation.  We repeated the same
calculation for $t$ = 5.10$^{-2}$, i.e. $T$ = 0.83~K, and we found $\theta$ = 30 degrees. 
However, at this temperature, we found a thinner film for which the
value of $\vartheta (x)$ is less accurately known.  It is
reasonable to find that the contact angle vanishes away from $T_{t}$
because the Casimir force vanishes while the van der Waals force
increases.  Clearly, there is a temperature region where $\theta$
increases with $T$, as found experimentally.  As for very close to
$T_{t}$, a crossover to a different regime should occur when 
$l_{eq} \approx \xi$ so that short range
forces should dominate; whether the contact angle keeps increasing, or reaches a finite 
value, or starts
decreasing to zero is an additional question to be solved.

Let us finally remark that, if we had an ordinary critical point with van 
der Waals forces and concentration fluctuations only, the Casimir force 
would be repulsive~\cite{Muk} and favor critical point 
wetting. In the case of our helium mixtures, 
it is the symmetric boundary conditions for superfluidity which lead 
to a 
Casimir force acting against critical point wetting.  One
obviously needs more measurements for a more precise determination of
$\theta$ and a calculation of the scaling function for a more accurate
theoretical prediction.

We are grateful to C. Guthmann, S. Moulinet, M. Poujade, D. Bonn, 
J. Meunier, D. Chatenay and E. Brezin for very helpful discussions. 
T. Ueno aknowledges support from the JSPS and the Kyoto University 
Foundation during his stay at ENS.

\end{multicols}
\end{document}